\documentclass[aps,reprint,amsmath,amssymb,eqsecnumaps,prr,superscriptaddress]{revtex4-2}
\usepackage{graphicx}% Include figure files
\usepackage{dcolumn}% Align table columns on decimal point
\usepackage{bm}% bold math
\usepackage{float}
\usepackage{xcolor}
\usepackage{lineno}
\newcommand{\vta}{\vartheta}
\newcommand{\p}{\partial}
\newcommand{\ep}{\varepsilon}

\newcommand{\vD}{\varDelta}

\newcommand{\om}{\omega}
\newcommand{\nn}{\nonumber}
\newcommand{\ta}{\theta}

\newcommand{\cH}{{\cal H}}
\newcommand{\cF}{{\cal F}}

\newcommand{\cW}{{\cal W}}
\newcommand{\ket}[1]{\vert{#1}\rangle}
\newcommand{\bra}[1]{\langle{#1}\vert}
\newcommand{\wh}{\widehat}
\newcommand{\wt}{\widetilde}
\newcommand{\be}{\begin{equation}}                                       
\newcommand{\ee}{\end{equation}}
\newcommand{\ba}{\begin{eqnarray}}
\newcommand{\ea}{\end{eqnarray}}
\newcommand{\bref}[1]{(\ref{#1})}
\newcommand{\lab}[1]{\label{#1}}

\newcommand{\bsub}{\begin{linenomath}\begin{subequations}}                      
\newcommand{\esub}{\end{subequations}\end{linenomath}}     
	
\begin{document}
	%	\setpagewiselinenumbers
	%\linenumbers
	\preprint{APS/123-QED}
	
	\title{Photon-photon polaritons in $\chi^{(2)}$ microresonators}% Force line breaks with \\
\author{D.V. Skryabin}   
\email{d.v.skryabin@bath.ac.uk} 
\affiliation{Department of Physics, University of Bath, Bath, BA2 7AY, UK}
\affiliation{Russian Quantum Centre, Skolkovo 121205, Russia}	
\author{V.V. Pankratov}   
\affiliation{Department of Physics, University of Bath, Bath, BA2 7AY, UK}
\author{A. Villois} 
\affiliation{Department of Physics, University of Bath, Bath, BA2 7AY, UK}
\author{D.N. Puzyrev}
\affiliation{Department of Physics, University of Bath, Bath, BA2 7AY, UK}
\date{Submitted 20 July 2020, revised 02 November 2020}

	\begin{abstract}
 We consider a high-Q microresonator with $\chi^{(2)}$ nonlinearity under conditions when the coupling rates between the sidebands around the pump and  second harmonic exceed the damping rates, implying the strong coupling regime (SC). Using the  dressed-resonator approach we demonstrate that this regime leads to the dominance of the Hermitian part of the operator driving the side-band dynamics over its non-Hermitian part 
responsible for the parametric gain.
This has allowed us to introduce and apply the cross-area concept of the  polariton 
quasi-particles and define their effective masses in the context of $\chi^{(2)}$ ring-microresonators. We further use polaritons to predict  the modified spectral response of the resonator to a weak probe field, and to reveal  splitting of the bare-resonator resonances,  avoided crossings, and Rabi dynamics. 
Polariton basis also allows deriving a discrete sequence 
of the parametric thresholds for the generation 
of sidebands of different orders.
	\end{abstract}
	\maketitle
%\date{\today}

\section{Introduction}
Strong coupling (SC) between photons and matter is at the heart of modern cavity quantum electrodynamics \cite{strong1}.  
Some of the currently burgeoning areas underpinned by the  SC physics are   quantum communication \cite{strong2}, microresonator exciton-polaritons \cite{sav,bog3,kav,sich,lau,im,bloch}, artificial  fields, atoms, and dimensions, and topological effects \cite{topo0,fan,topo}.

A common SC setup involves a mode of the high-finesse resonator interacting with a transition between the material energy levels. If the coupling rate is large relative to the dissipation rates, then the system can  sustain multiple  oscillations between the light and matter states~\cite{jc,cl,har}. The prime parameter behind properties of these oscillations is Rabi frequency, $\Omega$.  SC regime is most suitably described by the dressed eigenstates  hybridising  the light and matter degrees of freedom~\cite{dr,ram,scully,boyd0}. If the energy-momentum relation, $\ep(k)$, 
is introduced for the dressed states, then one can also define quasi-particles or elementary excitations~\cite{landau}. A list of hybrid light-matter quasi-particles, or polaritons,  includes exciton-polaritons~\cite{sav,bog3,kav,sich,lau}, plasmon-polaritons~\cite{bar,pl2}, EIT-polaritons (electromagnetically induced transparency)~\cite{lukin}, etc. Polaritons are not only a powerful theoretical concept, but are also a striking experimental feature  associated with 
splitting and avoided crossing  of the energy levels, 
and with Bose-Einstein condensation, see, e.g.,~\cite{sav,bog3,kav,pl2}.

While chip-integrated and bulk-cut high-Q ring microresonators 
continue to push limits of  frequency comb research~\cite{lipson,quant}, some experimental
results are also pointing towards potential these devices hold to study  SC.  
E.g., Ref.~\cite{tob} reported  SC between the whispering gallery modes and atoms,
Ref.~\cite{nori} studied the EIT-like regimes in two coupled resonators, and 
Ref.~\cite{gaeta} described measurements of the low-contrast resonance splitting effects
using  non-degenerate $\chi^{(3)}$ four-wave mixing (FWM). 
Refs.~\cite{falko,sipe2} developed theories for SC between two sidebands 
in $\chi^{(3)}$ FWM without exploring the possibility to introduce 
the energy-momentum relations and quasi-particles.

Monolithic- and microresonator devices with $\chi^{(2)}$ nonlinearity have 
been a viable and long-existing option complementing the  mainstream of 
$\chi^{(3)}$ work, see, e.g, \cite{mon1,mon2,rev3,prl0,prl1,prl2,ingo1}. 
Ref. \cite{car0} considered  SC between the pump and single-mode 2nd 
harmonic photons in a planar semiconductor cavity. 
$\chi^{(2)}$ response generally provides larger nonlinear phase shifts than  
$\chi^{(3)}$  \cite{rev3}, which is a factor increasing the coupling 
between sidebands. This is also complemented by the recent progress 
with simultaneous reduction of losses and volumes of $\chi^{(2)}$ microresonators  \cite{loncar,vahala,miro,bru,bru2,jan}. 
A combination of these factors favours considering if a multimode $\chi^{(2)}$ 
microresonator could go beyond simply modifying the light-matter interaction 
by reducing the density of states, i.e., Purcell regime, available for photon 
transitions and to cross into the SC regime, where new families of states (dressed states, 
and associated energy levels) are created.

Below, we demonstrate that  $\chi^{(2)}$ microresonators with
$Q$ approaching $10^8$ \cite{rev3,prl0,prl1,prl2,ingo1}, can operate in the SC regime 
between the sidebands centred around the pump (ordinary polarised) and 
second harmonic (extraordinary) frequencies. The SC regime
becomes accessible far from the resonance, well outside the bistability and soliton regimes \cite{bru2}. We parametrise energies of the dressed, i.e., hybrid ordinary-extraordinary, states 
with their momenta, and define microresonator quasi-particles -  {\em photon-photon polaritons}.  They are different from the above-mentioned families of the exciton-polaritons in planar  \cite{sav,bog3,kav,sich,lau,im,bloch} 
and ring \cite{egor} semiconductor microresonators, and from other polaritons 
involving transitions between the real, e.g,  molecular \cite{mol}, levels, because they do not require such absorbing transitions to exist in the matter component of the hybrid states. 
Instead, the matter side of the photon-photon polaritons belongs to the continuum of  virtual, i.e., far-away from resonances, electronic transitions. $\chi^{(2)}$ nonlinearity induced by these transitions is broad-band, and practically non-absorbing and non-dispersive.

\section{ Model}
The envelopes of the ordinary, $\psi_o$ (fundamental), and of the extraordinary, $\psi_e$ (2nd harmonic), fields in a microresonator are expressed via their mode expansions as $\psi_s=\sum_{\mu}\psi_{\mu s}(t)e^{i\mu\vta}$, with $s=o,e$ been the polarization state index.
$\vta\in[0,2\pi)$ is the polar angle varying along the 
resonator circumference, and
$\mu=0,\pm 1,\pm 2,\dots$ are the relative mode numbers, i.e., momenta. The associate resonance frequencies are $\om_{\mu s}=\om_{0 s}+\mu D_{1s}+\tfrac{1}{2}\mu^2 D_{2s}$, where $D_{1s}/2\pi$ are the repetition rates (free spectral ranges, FSRs) and 
$D_{2s}$ are the dispersions. 
Detunings of the resonator frequencies $\om_{\mu o}$ and $\om_{\mu e}$ from the pump laser frequency, $\om_p$,
and $2\om_p$ are  $\delta_{\mu o}=\om_{\mu o}-\om_p$
and $\delta_{\mu e}=\om_{\mu e}-2\om_p$, respectively.

Equations driving the envelope dynamics are~\cite{josab}
	\begin{align}
\nn
i\p_t\psi_o&=\delta_o\psi_o-iD_{1o}\p_\vta\psi_o-\tfrac{1}{2}D_{2o}\p^2_\vta\psi_o
\\ \nn  
&-\gamma_o\psi_e\psi_o^* -i\tfrac{1}{2}\kappa_o\big(\psi_o-\cH\big)
,\\ \nn
i\p_t\psi_e&=\delta_e\psi_e-iD_{1e}\p_\vta\psi_e-
\tfrac{1}{2}D_{2e}\p^2_\vta\psi_e
\\ 
&-\gamma_e\psi_o^2 -i\tfrac{1}{2}\kappa_e\psi_e.
\lab{main}
\end{align} 
Here $\delta_o\equiv\delta_{0 o}$, and $\delta_e\equiv\delta_{0 e}=2\delta_o-(2\om_{0o}-\om_{0e})$.
$\cH^2=\eta\cF\cW/\pi$ is the intra-resonator pump power, $\cW$ is the 'on-chip' laser power, $\cF=D_{1o}/\kappa_o\sim 10^4$ is finesse, and $\eta<1$ is the coupling efficiency.
$\kappa_s$ are the loaded linewidth parameters.
$\gamma_s$ are the nonlinear coefficients measured in Hz/W$^{1/2}$~\cite{josab}, see also~\cite{sm}
for more details. 
A Galilean transformation to the reference frame rotating with the 
rate $D_1/2\pi$, $\ta=\vta-D_1t$, converts detunings to 
\be
\vD_{\mu s}=\delta_{\mu s}-\mu D_{1}.
\lab{del0}\ee

\section{Dressed-resonator method} 
CW-solution of Eqs.~\bref{main}, $\p_\ta\wt\psi_{s}=\p_t\wt\psi_{ s}=0$, is sought  
in the form $\wt\psi_{o}/\cH_*=\Omega /\Omega_*$, where $\Omega_*=\sqrt{2\kappa_o\kappa_e}\simeq 
2\pi\times 6.3$MHz is the characteristic frequency scale defined by the linewidths,
and $\cH_*^2=\kappa_o\kappa_e/4\gamma_{2o}\gamma_{2e}\simeq 55\mu$W is the intra-resonator
power scale.  
Thus, $\Omega$ is the frequency measure of the intra-resonator field, and, as we will see below, it has meaning of the Rabi frequency.

CW-state power
is now expressed via modulus of $\Omega$, $|\wt\psi_{ o}|^2=\cH_*^2|\Omega|^2 /\Omega_*^2$,  and straightforward algebra with Eqs.~\bref{main} reveals that $\Omega$ is a solution of 
\be
i\frac{\Omega}{\Omega_*}\cdot\frac{\Omega_o}{\kappa_o/2}\left(
1-\frac{|\Omega|^2}{\Omega_o\Omega_e}\right)=\frac{\cH}{\cH_*}=\sqrt{\frac{\cW}{\cW_*}},
\lab{cw}\ee
and that the 2nd harmonic amplitude is $\wt\psi_{ e}=\Omega^2/\Omega_e\gamma_o$. 
Here  $\cW_*=\pi\cH_*^2/\eta\cF\simeq 16$nW is the 'on-chip' pump power scale, and 
$\Omega_o\equiv(\delta_o-i\kappa_o/2)$,  
$\Omega_e\equiv 8(\delta_e-i\kappa_e/2)$ are the complex frequencies introduced 
for the sake of brevity.

Following the  ethos of the dressed-atom theory \cite{dr,ram,scully,boyd0}, 
we  apply the dressed-resonator method consisting from (i)~expanding the small 
amplitude perturbations around the cw-state using the modes of the linear resonator, (ii)~separating the Hermitian and non-Hermitian parts
of the operator driving the evolution of the perturbations, and (iii)~identifying the eigenstates of the Hermitian part, i.e., dressed-states or polaritons.  
\begin{figure*}[t]
	\centering
	\includegraphics[width=0.9\textwidth]{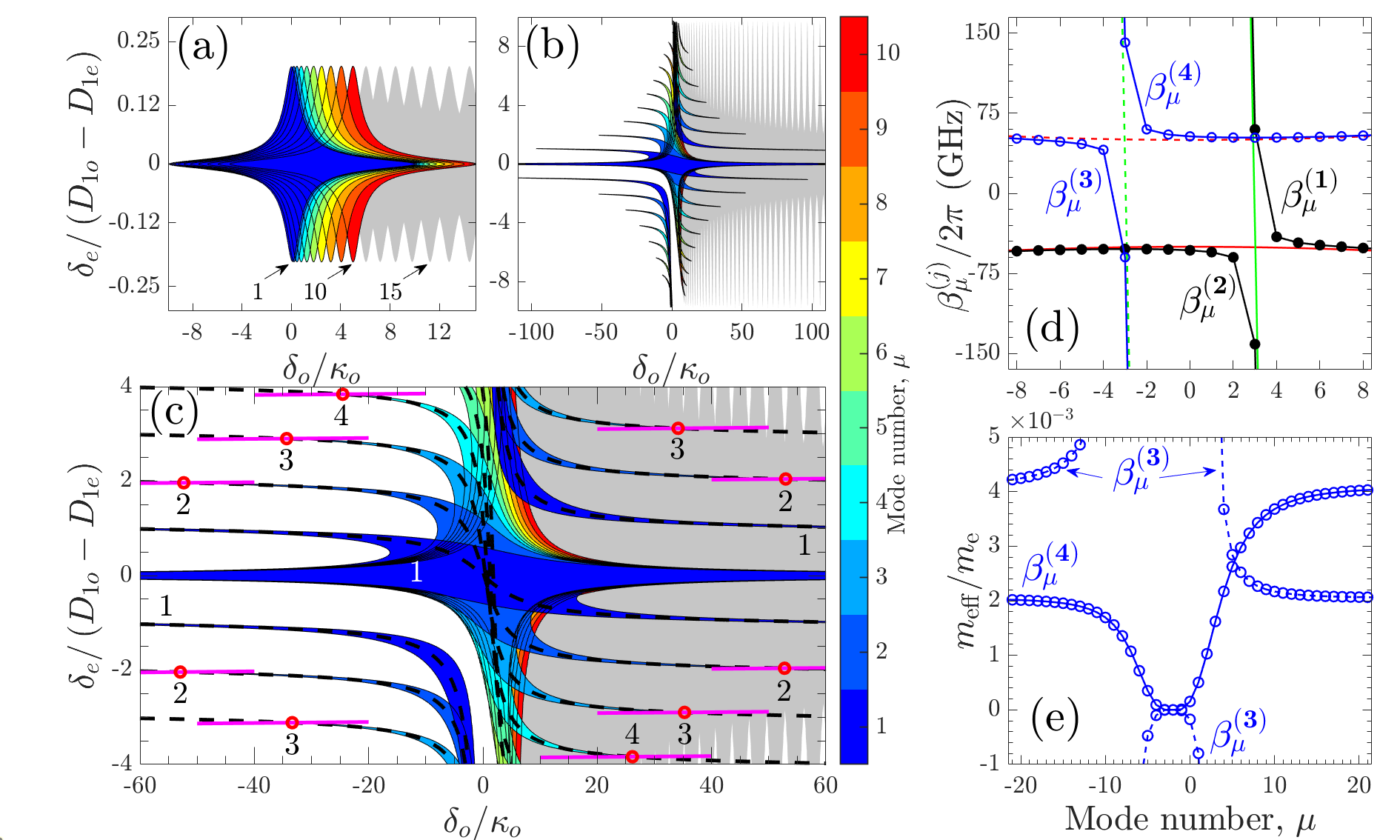}
	\caption{{\bf (a-b)}~Reshaping of the $\mu\ne 0$ parametric instability 
		boundaries  for $|\Omega|$ increasing from $28\times \kappa_o$ (a) to 
		$200\times \kappa_o$ (b). Coloured and grey shading cover the unstable domains.
		{\bf (c)} shows the finer details of (b).  Red points mark the tips of the unstable tongues, and the numbers are the respective $\mu$'s.  Dashed black lines show the resonance condition, Eq. \bref{res}, and the short magenta lines show the threshold conditions $V_\mu^{(j_1j_2)}V_\mu^{(j_2j_1)}=V_\mu^{(j_1j_1)}V_\mu^{(j_2j_2)}$;  
		$j_1,j_2=2,4$ in the 1st quadrant;
		$1,4$ in the 2nd and 4th; $1,3$ in the 3rd.
		Other parameters are   $(D_{1o}-D_{1e})/\kappa_o=10^3$, $D_{2o}/\kappa_o=-0.1$, $D_{2e}/\kappa_o=-0.2$. The colorbar shows the correspondence between the colors and the mode-numbers, $|\mu|$.
		{\bf (d)}~The photon-photon frequency $\beta^{(j)}_\mu$ vs momentum  $\mu$:  
	 $\delta_o/\kappa_o=-50$, $\delta_e/(D_{1o}-D_{1e})=2.97$, are taken close  to the  $\mu=3$ tip in (c).  Red and green lines without circles mark the bare, $\Omega=0$, spectrum of $\wh H_\mu$.
	The avoided crossings  are centred
	at $\vD_\mu=0$ and $\vD_{-\mu}=0$. {\bf (e)}~Effective photon-photon mass, $m_{\text{eff}}$, relative to the electron mass, $m_{\text{e}}$, for the $\beta^{(3)}_\mu$ (dashed line), and $\beta^{(4)}_\mu$ (full line) branches from (d). }
	\label{f1}
\end{figure*}

We now look for a solution of Eqs.~\bref{main} in the form, 
$\psi_s=\wt\psi_{s}+q_se^{i\phi_{s}}
\textstyle{\sum_{\mu}}\big(a_{\mu s}(t)e^{i\mu\ta}+
a^*_{-\mu s}(t)e^{-i\mu\ta}\big)$,
where $q_o=1/\sqrt{2}$,  $q_e=\sqrt{\gamma_e/\gamma_o}$ and $\phi_{ s}=\text{arg}~\wt\psi_{s}$.
Assuming that the sideband amplitudes, $|a_{\mu s}|$, are small we find
\be
i\p_t\ket{a_\mu}=\big(\wh H_\mu+\wh V\big)\ket{a_\mu}.
\lab{master}
\ee
Here  $\ket{a_\mu}=(a_{\mu o},a_{\mu e},a_{-\mu o},a_{-\mu e})^T$ is
the state vector, 
\begin{align}
\wh H_\mu=\left[\begin{array}{cccc}
\vD_{\mu o} 
&-\tfrac{1}{2}|\Omega| e^{-i\phi}
& 0
&0
\\
-\tfrac{1}{2}|\Omega| e^{i\phi}
& \vD_{\mu e} 
& 0 
& 0
\\
0
& 0 
& -\vD_{-\mu o} 
&\tfrac{1}{2}|\Omega| e^{i\phi}
\\
0 
& 0
& \tfrac{1}{2}|\Omega| e^{-i\phi} 
&-\vD_{-\mu e}
\end{array}\right],
\lab{ham}
\end{align}
$\phi=2\phi_{o}-\phi_{e}$, and 
\begin{align}
\wh V=\left[\begin{array}{cccc}
-i\tfrac{1}{2}\kappa_o 
&0
& -\tfrac{|\Omega|^2}{|\Omega_e|}e^{-i\phi} 
&0
\\
0
&  -i\tfrac{1}{2}\kappa_e
& 0 
& 0
\\
\tfrac{|\Omega|^2}{|\Omega_e|}e^{i\phi} 
& 0 
& -i\tfrac{1}{2}\kappa_o
&0
\\
0 
&0
& 0 
&-i\tfrac{1}{2}\kappa_e
\end{array}\right],
\lab{v}
\end{align}

The diagonal terms in $\wh H_\mu$ make up the Hamiltonian of the bare, i.e., undressed, resonator. 
$\Omega\ne 0$ provides dressing  and, importantly, retains the Hermitian structure.
The diagonal terms in $\wh H_\mu$ are $\sim |\Omega|^1$, and drive the flopping between 
the bare ordinary and extraordinary states with the same $\mu$, see the state-vector structure.
Therefore, $\Omega$ has the meaning of the Rabi frequency.
Like in the two-level atom case \cite{scully,boyd0}, the bare  states are separated 
by the optical frequency, i.e., by the energy $\sim$eV. While, the level splitting is 
characterised by the radio-frequency (RF) scale, $\hbar|\Omega|\lesssim 1\mu$eV~\cite{sm}.
The off-diagonal terms in $\wh H_\mu$ are the three-wave mixing (TWM) ones, since they describe adding one $\mu=0$ ordinary photon, defined by the $|\Omega|^1$-term, $\Omega\sim\wt\psi_o$,  to one ordinary $\mu\ne 0$ photon to generate one extraordinary photon with the same $\mu$.
The two-level atom methodology was also previously developed for the resonator-free 
sum-frequency generation, see, e.g., \cite{yaron,ady}.

Interaction between the $\mu$ and $-\mu$   ordinary photons is governed by the off-diagonal  terms in $\wh V$. These are the four-wave mixing (FWM) terms engaging two $\mu=0$ ordinary photons, coming from $|\Omega|^2$, and  two side-band ones, which are also ordinary, but have momenta $\pm\mu$. For studies into the FWM gain in  $\chi^{(2)}$ materials see, e.g., Refs.~\cite{rev3,steg}.
$\wh V$ also includes dissipation, that competes with the FWM gain. 
Therefore, the sideband dynamics is expected to be dominated by the TWM Rabi dynamics for
$||\wh H_\mu||\gg ||\wh V||$. This constitutes the SC conditions, which can be expressed 
as $|\Omega_e|\gg |\Omega|\gg\Omega_*$. $|\Omega|\gg\Omega_*\equiv\sqrt{2\kappa_o\kappa_e}$ 
is achieved by taking a high-Q resonator and the relatively strong pump, while $|\Omega|\ll |\Omega_e|$ implies small conversion efficiency to the 2nd harmonic due to large value of the frequency mismatch parameter, 
$|2\om_{0o}-\om_{0e}|\sim |\Omega_e|/8$. Note, that Eq.~\bref{cw} can be linearised for $|\Omega|^2/|\Omega_e||\Omega_o|\ll 1$, and then the Rabi frequency is estimated as $|\Omega|\approx\Omega_* (\kappa_o/2|\delta_o|)\sqrt{\cW/\cW_*}$~\cite{sm}.

\section{Results} Before using SC conditions to  our advantage, 
we set $\ket{a_\mu}=\ket{b_\mu} e^{-i\beta_\mu t}$ and  solve the full 
dispersion law $\text{det}(\wh H_\mu+\wh V-\beta_\mu\wh I)=0$ numerically 
for every $\mu$. Condition $\text{Im}\beta_\mu=0$ specifies the boundaries 
between the areas with and without gain. Fig.~\ref{f1}(a) shows the case of $|\Omega|/\kappa_o\gg 1$, 
and $|\Omega|/\kappa_e\sim 1$, i.e., when we have SC for the ordinary, but not yet for the extraordinary photons. Under these conditions, the $\mu$-specific instability domains 
are arranged as a sequence of the 4-beam stars shifted along the $\delta_o$, 
but nearly coinciding along the $\delta_e$, direction.  
As $|\Omega|$ is increased further and provides $|\Omega|/\kappa_{o,e}\gg 1$, i.e., 
$|\Omega|\gg\Omega_*$,
we step firmly into the SC regime. Now, a spectacular pattern of the narrow 
resonances appears along the o $\delta_e$ direction and the ones along $\delta_o$ 
sharpen even more, see Figs.~\ref{f1}(b),(c).   

Thus, the SC condition channels gain into the narrow mode-number specific regions 
in the parameter space. 
Parametric frequency conversion  in $\chi^{(2)}$ resonators with $\cF\sim 10$-$10^2$, 
operating in the weak-coupling, i.e., Purcell, regime, that has  attracted significant recent attention~\cite{ent,ulv,wab1,wab2}, does not have this effect, and the contrast of 
the star beams in Fig.~\ref{f1}(a) is reduced further for smaller $\cF$.
Formation of the narrow instability tongues in the high-$\cF$ regime is similar to the 
Arnold-tongues reported  in the high-finesse Kerr microresonators~\cite{arnold,pra}. However, the matrix operator  driving the sideband dynamics  in 
Refs.~\cite{arnold,pra} is entirely non-Hermitian, as is $\wh V$ here. Therefore, it does not allow to define the energy and momentum for the quasi-particles in the way possible here, if 
$\wh H_\mu$ dominates over $\wh V$, i.e., in the SC regime.
 
To leading order, the mode dynamics in the SC regime is driven by 
$i\p_t\ket{a_\mu}=\wh H_\mu\ket{a_\mu}$.  $\wh H_\mu$ is block-diagonal
and has two orthogonal sub-spaces of the dressed states. The top-block subspace 
is span by $\ket{b_\mu^{(1)}}=\big(|\Omega| e^{-i\phi},\vD_{\mu}-\Omega_\mu,0,0\big)^T$ 
and $\ket{b_\mu ^{(2)}}=\big(\Omega_\mu-\vD_{\mu},|\Omega| e^{i\phi},0,0\big)^T$, where 
\be
\vD_\mu=
\vD_{\mu o}-\vD_{\mu e}, 
\lab{off}
\ee
is the Rabi detuning, see Eqs.~\bref{bog}.
The corresponding eigenfrequencies are 
$\beta_\mu^{(1)}=\beta_\mu^{(+)}$ and $\beta_\mu^{(2)}=\beta_\mu^{(-)}$, 
\be
\beta_\mu^{(\pm)} =\tfrac{1}{2}(\vD_{\mu o}+\vD_{\mu e})\pm\tfrac{1}{2}\Omega_\mu,~
\Omega_\mu=
\sqrt{\vD_{\mu 
		}^2+|\Omega|^2}.
\lab{bog}
\ee

\begin{figure*}[t]
\centering
\includegraphics[width=0.8\textwidth]{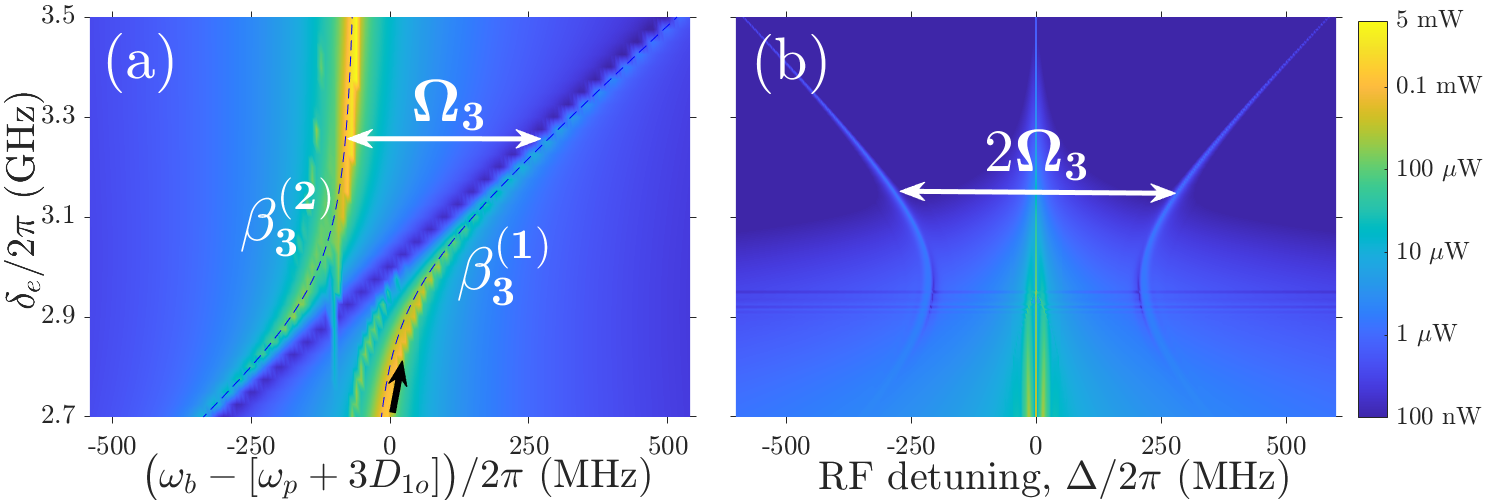}
\caption{		
{\bf (a)}~Avoided crossing of the $\beta_3^{(1)}$ and $\beta_3^{(2)}$ 
photon-photon states as recorded through the series of the numerical scans of the probe frequency, $\om_b$, 
for a range of $\delta_e$ values. $\om_b$ is scanned around the bare resonator resonance at $\om_{3o}\simeq \om_{0 o}+3D_{1o}$. Color density shows the $\mu=3$ mode power, $\int_0^\tau |\psi_{3o}|^2dt/\tau$, see color-bar. Parameters are  $\cW_b=0.9$mW (probe laser power), $\cW\simeq 176$mW (pump laser power), $\delta_o/2\pi=-50$MHz, $\tau =0.4\mu$s. Corresponding cw-state power, i.e., the one that directly drives nonlinear 
processes in the resonator material, is  $|\wt \psi_{o}|^2=\cH_*^2|\Omega|^2/\Omega_*^2\simeq (\eta\cF\cW/4\pi)\times (\kappa_o/\delta_o)^2\simeq 60$mW, which is below $\cW$ because the resonator is driven far off-resonance.
		{\bf (b)}~Rabi (Autler-Townes) sidebands in the RF spectrum, $\int_0^\tau |\psi_{3o}|^2e^{-it\Delta }dt/\tau$ vs $\delta_e$, as computed for the $\beta_3^{(1)}$ polariton branch, 
		see the black arrow in~(a). Color-bar applies to both (a) and (b).
}
	\label{f2}
\end{figure*}

Here, $\Omega_\mu$ is the effective Rabi frequency~\cite{boyd0,scully}.
The bottom-block eigenstates and eigenfrequencies are
$\ket{b_{\mu}^{(3)}}=\big(0,0,|\Omega| e^{i\phi},\vD_{-\mu}-\Omega_{-\mu}\big)^T$, 
$\beta_\mu^{(3)}=-\beta_{-\mu}^{(+)}$, and 
$\ket{b_{\mu}^{(4)}}=\big(0,0,\Omega_{-\mu}-\vD_{-\mu},|\Omega| e^{-i\phi}\big)^T$,
$\beta_\mu^{(4)}=-\beta_{-\mu}^{(-)}$. 
Each pair of the dressed states makes its own avoided crossing around the momentum 
values corresponding to $\vD_{\pm\mu}=0$,  Fig.~\ref{f2}(a). 
Close to the avoided crossings, the state dressing is increasing via  $\Omega_{\pm\mu}-\vD_{\pm\mu}\to|\Omega|$.

The  frequency-momentum dependencies, $\beta^{(j)}_\mu$ vs $\mu$ for $j=1,2$ and $j=3,4$, 
Fig. \ref{f1}(d), define the four families of quasi-particles. These quasi-particles hybridise the $\om_{\mu o}$  and $\om_{\mu e}$  photons, and can be naturally called  - {\em photon-photon polaritons}.  
$\wh H_\mu$ is Hermitian and its eigenfrequencies can be used as the 
quasi-particle energy spectrum. Hence,
the photon-photons carry energy $\ep=\hbar\beta^{(j)}_\mu$ ($\sim\mu$eV, $D_1\sim D_{1o}$), angular momentum $\ell=\hbar\mu$, and the linear momentum $k=\ell/R$, where $R$ is the resonator radius.  
Predictably, there is a flexibility with defining the  photon-photon zero-energy level. This is controlled by the reference frame choice, i.e., by $D_1$ entering $\vD_{\mu o}+\vD_{\mu e}$, while $\vD_\mu$ and $\Omega_\mu$ are $D_1$ independent. 
The effective photon-photon mass  is evaluated as 
$m_{\text{eff}}^{-1}=\p_k^2[\hbar\beta_\mu^{(j)}]$. Away from the avoided crossings, 
the photon-photons behave as photons and have  $m_{\text{eff}}\simeq \pm\hbar/R^2|D_{2s}|\sim 10^{-3}m_{\text{e}}$, where
$m_{\text{e}}$ is the electron mass. Close to the avoided crossings, SC modifies linear 
dispersion, and, because of this, $m_{\text{eff}}$ can drop by few orders of magnitude, 
sharply rise to infinity 
and change its sign, see Fig.~\ref{f1}(e).

Photon-photons make the four spectral families, when their frequencies are recalculated back to the physical spectral range, 
$\widetilde\omega_{\mu o}^{(j)} = 
\omega_p\pm\beta_\mu^{(j)}\pm\mu D_1$ and 
$\widetilde\omega_{\mu e}^{(j)} = 
2\omega_p\pm\beta_\mu^{(j)}\pm\mu D_1$, where '$+$' should be used for $j=1,2$ and '$-$' for $j=3,4$.
While these dressed spectra are not equidistant, they provide equal separation of the sideband lines from the pump and its second harmonic, i.e.,
$\wt\omega_{\mu o}^{(j)}-\wt\omega_{-\mu o}^{(j)}=\wt\omega_{\mu e}^{(j)}-\wt\omega_{-\mu e}^{(j)}$.   
This is quite a remarkable observation, since  for the bare-photons  $\big((\om_{\mu o}-\om_{-\mu o})-(\om_{\mu e}-\om_{-\mu e})\big)\sim (D_{1o}-D_{1e})|\mu|$ reaches
to GHz. 

Let us now describe how the dressed spectra allow formulating the TWM and FWM energy conservation laws.  
TWM drives the Rabi flops between $|\psi_{\mu o}(t)|^2$ and $|\psi_{\mu e}(t)|^2$, and the quanta involved satisfy 
\be\hbar\om_p+\hbar\wt\om^{(j)}_{\mu o}=\hbar\wt\om_{\mu e}^{(j)}.\nn\ee
The above condition combines a pump photon with a
dressed ordinary photon with momentum $\mu$ to produce 
a dressed extraordinary photon with the same $\mu$. 
The other process, is the cross-branch FWM that engages $\mu$ and $-\mu$ quanta 
and  yields the following energy conservation between two pump laser photons 
and two dressed ordinary polarised quanta,
\be
\hbar\om_p+\hbar\om_p=
\hbar\wt\om_{\mu o}^{(j_1)}+\hbar\wt\om_{\mu o}^{(j_2)},~j_1=1,2,~j_2=3,4.
\lab{en}\ee

An important feature of the $\beta_\mu^{(j)}$ vs $\mu$ dependencies 
are the resonances between the  photon-photons 
belonging to the $1,2$ (full circles in Fig.~\ref{f1}(d)) and $3,4$ (empty circles) branches. 
Four eigenfrequencies generally admit
six distinct resonance conditions, $\beta_{\mu }^{(j_1)}=\beta_{\mu }^{(j_2)}$. 
However, all six of them can be realised only for $\Omega=0$. 
The dressed resonator, $\Omega\ne 0$,  allows for the four resonance conditions
$(j_1;j_2)=(1;3)$, $(1;4)$, $(2;3)$, and $(2;4)$, each yielding the photon energy conservation in Eq.~\bref{en}. 
A particular case selected for  Fig.~\ref{f1}(d) illustrates two simultaneous resonance conditions
$\beta_\mu^{(1)}=\beta_\mu^{(4)}$ and  $\beta_{\mu}^{(2)}=\beta_{\mu}^{(3)}$. 
The $(j_1;j_2)=(1;2)$ and $(3;4)$ resonances existing for $\Omega=0$ are replaced with 
the avoided crossings for $\Omega\ne 0$.

Resolving each of the four resonance conditions for $\Omega$ gives the same elegant answer,
\begin{align}
\frac{|\Omega|^2}{4}
&=
\frac{
	(\vD_{\mu o}+\vD_{-\mu e})
	(\vD_{\mu o}+\vD_{-\mu o}) 
	(\vD_{\mu e}+\vD_{-\mu e})
}
{(
	\vD_{\mu o}+\vD_{-\mu o}+
	\vD_{\mu e}+\vD_{-\mu e})^2}\nn\\ 
& \times
(\vD_{\mu e}+\vD_{-\mu o}).
\lab{res}
\end{align}
Eq.~\bref{res} is structurally similar to the resonant denominators derived for 
the  susceptibilities emerging through the interaction of the multimode fields with the multilevel 
atoms~\cite{boyd0,scully,dr}. The resonance lines computed from Eq.~\bref{res} and mapped onto Fig.~\ref{f1}(c) precisely follow the maximal gain lines  in the parameter space, i.e., $\max\text{Im}\beta_\mu>0$ vs $\delta_s$, where the complex $\beta_\mu$ are computed from $\text{det}(\wh H_\mu+\wh V-\beta_\mu\wh I)=0$.
Taking Fig.~\ref{f1}(c), we find that the first and last brackets in Eq.~\bref{res} determine the resonance separation along  $\delta_e$, which  happens in steps  $\simeq |D_{1o}-D_{1e}|$,
2nd bracket controls resonances along  $\delta_o$ in steps of $\simeq |D_{2o}\mu|$, while the 3rd bracket does not make a noticeable impact along $\delta_e$ because $|D_{2e}\mu|\ll |D_{1o}-D_{1e}|$.

All resonance lines have a special point  triggering a  parametric instability relative to the $\pm\mu$ modes.
To find the parametric gain threshold when the system is confined to one of the resonances, i.e., when two energy levels are degenerate,  $\beta_\mu^{(j_1)}=\beta_\mu^{(j_2)}$, while their eigenvectors remain different,
we apply the degenerate state  perturbation theory~\cite{fermi} to Eq.~\bref{master}  by treating the FWM operator $\wh V$ as a perturbation. The matrix elements of $\wh V$ are $\bra{b_\mu^{(j_1)}}\wh V\ket{b_\mu^{(j_2)}}=V_\mu^{(j_1j_2)}$.
FWM removes the degeneracy between  $\beta_\mu^{(j_1)}$ and $\beta_\mu^{(j_2)}$ 
by  making different the loss vs gain balance acquired by the two states. One state becomes lossy, while the other generates gain providing
$V_\mu^{(j_1j_2)}V_\mu^{(j_2j_1)}>V_\mu^{(j_1j_1)}V_\mu^{(j_2j_2)}$. 
Opening the above for, e.g., the $(j_1;j_2)=(1;4)$ case yields  
\begin{align}
&
\frac{|\Omega|^6}{|\Omega_e|^2}\big(\vD_{-\mu}-\Omega_{-\mu}\big)^2=\frac{1}{4}\times
\lab{tan}
\\
\nn &\big(\kappa_o|\Omega|^2+\kappa_e(\vD_\mu-\Omega_\mu)^2\big)
\big(\kappa_e|\Omega|^2+\kappa_o(\vD_{-\mu}-\Omega_{-\mu})^2\big).
\end{align}
Eq.~\bref{tan} explicitly expresses the balance between the net gain (left) and the net loss (right).
The intersection between the hypersurfaces defined by Eqs.~\bref{res} and \bref{tan} provides an excellent approximation for the locations of the instability boundary tips, see Fig.~\ref{f1}(c), thereby proving that the resonance effect lies at the origin of the channelling of the parametric gain into narrow instability domains.

Dressing of the spectrum, i.e., splitting of the resonances of the bare, $\Omega=0$, resonator, and 
the underpinning Rabi dynamics have been also demonstrated by us in a series of numerical experiments reproducing how the photon-photon realm could be measured in a lab. 
To achieve this, we  consider a region in Fig.~\ref{f1}(c) that avoids the FWM gain, and where the power enhancement effects provided by the resonator are diminished or even reversed, while the SC induced dressing effects on the resonator spectrum can be seen clearly. 
We then include a weak probe field, '$b$', into Eqs.~\bref{main},  through the transformation $\cH\to\cH+\cH_b e^{i\mu\vta-i(\om_b-\om_p)t}$. 
This  probe field has the non-zero projections on the $\ket{b_\mu^{(1)}}$, $\ket{b_\mu^{(2)}}$  
dressed states  via their ordinary components.  Therefore, if $\om_b$ is tuned to the frequency  of one of the  dressed states, e.g., to $\wt\om_{\mu o}^{(1)}$,  it then resonantly excites a superposition state,
$\alpha_1\ket{b_\mu^{(1)}}e^{-it\beta_\mu^{(1)}}+\alpha_2\ket{b_\mu^{(2)}}e^{-it\beta_\mu^{(2)}}$, that exhibits   Rabi oscillations. These oscillations are expressed as the anti-phase beats of $|\psi_{\mu o}(t)|^2$ and $|\psi_{\mu e}(t)|^2$ 
with the frequency $\wt\om_{\mu s}^{(1)}-\wt\om_{\mu s}^{(2)}=\beta_{\mu}^{(1)}-\beta_{\mu}^{(2)}=\Omega_\mu$.
The probe laser power, and what it achieves inside the resonator during the frequency scan, is relatively weak, 
while the pump laser is kept at the constant frequency far away from the resonance,  and therefore they 
are not expected to cause significant thermal effects,  
see Fig.~\ref{f2}  for the power and detuning values.  Other pump-probe arrangements in Kerr microresonators have been 
recently demonstrated in Ref.~\cite{pasc}. Further thermal control 
can be applied with the techniques as the ones developed for cooling of 
microresonators and photonic chips~\cite{papp1,cr4,cg1}.

Fig.~\ref{f2}(a) shows how the  $|\psi_{3 o}|^2$ power responds 
to the applied probe depending on its frequency, $\om_b$, and through a range of the $\delta_e$ values. One can see pronounced resonant responses for $\om_b$  tuned to $\wt\om_{3  o}^{(1)}$ and $\wt\om_{3 o}^{(2)}$. This fully confirms our analytical results for the dressed spectrum, see Eqs. \bref{bog} and the dashed lines in Fig. \ref{f2}(a). The avoided crossing region in Fig.~\ref{f2}(a) corresponds to the minimum of $\Omega_3$  in $\delta_e$, i.e., $\vD_3=0$ and $\Omega_3=|\Omega|$. 
Same avoided crossings can be observed for the whole sequence of 
the resonances in the 'star'-diagram in Fig.~\ref{f1}(c). 
Measurements like in Fig.~\ref{f2}(a) are often used as signatures of 
the polariton existence in other optical systems, see, e.g., Refs.~\cite{kav,bar}.

Rabi beats can be observed by 
measuring the RF spectra  of either the total field or 
the individual modes, $\mu=3$ in this case.
Fig.~\ref{f2}(b) shows the spectrum of $|\psi_{3o}(t)|^2$ 
vs $\delta_e$ with $\om_b$ tuned to $\wt\om_{3 o}^{(1)}$.
It makes obvious the spectral signature of the Rabi splitting, which is also known 
 as the Autler-Townes splitting in  spectroscopy \cite{dr,nori,cr3}. 
The sideband power in Fig.~\ref{f2}(b) is proportional to the modulation depth of $|\psi_{3o}(t)|^2$,
which is reduced if the probe projection on one of the two dressed states 
is much larger than on the other, e.g., $|\alpha_1|\gg|\alpha_2|$.  

\section{Summary} We have proposed and formalised a concept of the polariton quasi-particles, photon-photon polaritons, in the context of multimode high-Q $\chi^{(2)}$ microring resonators. The quasi-particle regime becomes possible if a resonator is pumped far-off the resonance,  which allows the dominance of the Hermitian part of the side-band coupling operator over the dissipation and parametric gain effects. Such a resonator operates in the strong-coupling regime, shows the splitting of the resonances, their avoided crossings, and Rabi oscillations between the ordinary and extraordinary modes. Our results reveal a promising connection between the rapidly expanding  research area of frequency conversion in high-Q microresonators and the quasiparticle approach widely used in 
other branches of photonics and condensed matter physics.

\begin{acknowledgments}
	This work was supported by the EU  Horizon 2020 
	Framework Programme (812818, MICROCOMB), UK EPSRC (EP/R008159),
	and Russian Science Foundation (17-12-01413-$\Pi$).

\end{acknowledgments}

\end{document}